\documentclass[conference]{IEEEtran}
\IEEEoverridecommandlockouts
\usepackage{cite}
\usepackage{amsmath,amssymb,amsfonts,amsthm}
\usepackage{algorithmic}
\usepackage{graphicx}
\usepackage{textcomp}
\usepackage{xcolor}
\usepackage{wasysym}
\usepackage{newtxtext,newtxmath, array, mathrsfs,multirow}
\usepackage[caption=false]{subfig}
\usepackage{stfloats}
\usepackage{threeparttable}
\usepackage{cases}
\usepackage[linesnumbered, ruled]{algorithm2e}
\SetKwRepeat{Do}{do}{while}%
\usepackage{cite}
\usepackage[all=normal, paragraphs=tight, floats=tight, mathspacing=tight]{savetrees}

\usepackage{comment} 

\usepackage{bm}

\begin{document}

\title{Null-Shaping for Interference Mitigation in\\LEO Satellites Under Location Uncertainty}
\vspace{-0.5cm}
\author{ 
\IEEEauthorblockN{
Fernando Moya Caceres,
Akram Al-Hourani,
Saman Atapattu, and
Sithamparanathan Kandeepan
}
 \IEEEauthorblockA{
 Department of Electrical and Electronic Engineering, School of Engineering, RMIT University, Melbourne,  Australia. \\
\IEEEauthorblockA{Email:
\{fernando.moyacaceres, akram.hourani,  saman.atapattu, kandeepan.sithamparanathan\}@rmit.edu.au
}
}
\vspace{-1.0cm}
}
\maketitle
\begin{abstract}
Radio frequency interference (RFI) poses a growing challenge to satellite communications, particularly in uplink channels of Low Earth Orbit (LEO) systems, due to increasing spectrum congestion and uncertainty in the location of terrestrial interferers. This paper addresses the impact of RFI source position uncertainty on beamforming-based interference mitigation. First, we analytically characterize how geographic uncertainty in RFI location translates into angular deviation as observed from the satellite. Building on this, we propose a robust null-shaping framework to increase resilience in the communication links by incorporating the probability density function (PDF) of the RFI location uncertainty into the beamforming design via stochastic optimization. This allows adaptive shaping of the antenna array's nulling pattern to enhance interference suppression under uncertainty. Extensive Monte Carlo simulations, incorporating realistic satellite orbital dynamics and various RFI scenarios, demonstrate that the proposed approach achieves significantly improved mitigation performance compared to conventional deterministic designs. 
\end{abstract}


\section{Introduction}
Evolution in Satellite Communications (SatCom) systems is motivated by the need to provide ubiquitous and uninterrupted service around the globe. In particular, Low Earth Orbit (LEO) satellite constellations, are undergoing substantial expansion due to important advantages such as improved propagation delay, reduced deployment cost \cite{8700141}, and the emergence of new technologies such as satellite links. Recent directions include the attempt to integrate satellite and terrestrial networks which is largely motivated as part of next generation cellular system~\cite{9509510,10045716}, the support of emerging applications namely Internet-of-Things (IoT)~\cite{8002583}, and localization~\cite{9810526}. Networks including Starlink, Eutelsat-OneWeb, and Kuiper are actively expanding. Due to the large number of satellite, such constellations are dubbed ``mega-constellations''~\cite{10286330,9313025,Guo2025vt} having launched more than 7800 satellites in the last 8 years~\cite{10209551}, and with many more planned for the upcoming years~\cite{9755278}, which shows the significant interest in exploring LEO networks. As with any thriving technology, rapidly attracting a great number of users and services, reliability becomes a mandatory aspect to take into account. Among all threats that LEO networks are exposed to, we dive into the prominent {\it interference} issues that result from either intentional or unintentional sources, originating from both terrestrial and space systems~\cite{10209551}. When intentionally generated, jamming is a simple yet effective strategy capable of disrupting communication links~\cite{9733393}. 

In SatCom, when addressing the Radio frequency interference (RFI) issue, beamforming-based mitigation is a powerful technique to cope with unwanted signals by reducing their impact directly in the physical layer and enhances overall system resilience~\cite{10622540}. Its effectiveness is associated with possessing precise information on the direction of the RFI source(s). When beamforming design follows this assumption, nulls can be carefully generated to reduce the received power from those specific locations. Usually, when analytic methods are used, the resulting null-widths are extremely narrow in both the spatial domain and frequency domain~\cite{s22186984}. In practice, obtaining high precision locations of RFI sources is problematic due to many factors including, platform attitude, accuracy of the localization algorithms, and RFI sources movement~\cite{9531137,10622540,s22186984}. 

The uncertainty in the RFI source location significantly impacts the mitigation performance as RFI sources do not perfectly match the designed {\it nulling space}~\cite{6511365}. Consequently, when ambiguity is present, mitigation performance effectiveness decays abruptly~\cite{10622540}. An alternative to enhance beamforming interference mitigation under these circumstances is to shift efforts to properly adapt the nulling angular width considering uncertainties in the available location information. For example, research on shaping the null-width in linear and planar arrays through the presence of virtual interference~\cite{9531137,s22186984} has shown to be effective. Due to the complexity of finding an optimal analytic solution for large antenna arrays, the use of evolutionary algorithms~\cite{4016795,8669115} has been proposed as an alternative to deal with such a demanding task.

To the best of the authors’ knowledge, the adaptation of an antenna array’s nulling space to account for uncertainty in the location of RFI sources has not been previously addressed in the literature. {\it This work focuses on uplink interference mitigation for LEO satellite communication links in the presence of RFI sources with uncertain positions. We propose a beamforming-based nulling framework that leverages statistical knowledge of RFI location uncertainty to enhance interference suppression}. The main contributions of this paper are:
\begin{enumerate}
    \item A geometrical analysis of the angular deviation between the expected and actual RFI source positions as observed from the satellite, and its corresponding orthodromic distance on Earth. We quantify its impact on nulling effectiveness and provide design insights for parameter selection. 
    \item A novel interference mitigation framework that shapes the beamforming nulls by incorporating statistical information on RFI source position uncertainty, thereby improving robustness in practical LEO scenarios. 
    \item An adaptive nulling strategy based on a weighted optimization approach, where the probability density function (PDF) of the RFI location uncertainty is used to dynamically adjust the antenna array’s radiation pattern for enhanced interference suppression.
\end{enumerate}

\section{System Model}\label{Sec_SysModel}
\begin{figure}[t!]
    \centering
    \includegraphics[width=0.5\textwidth]{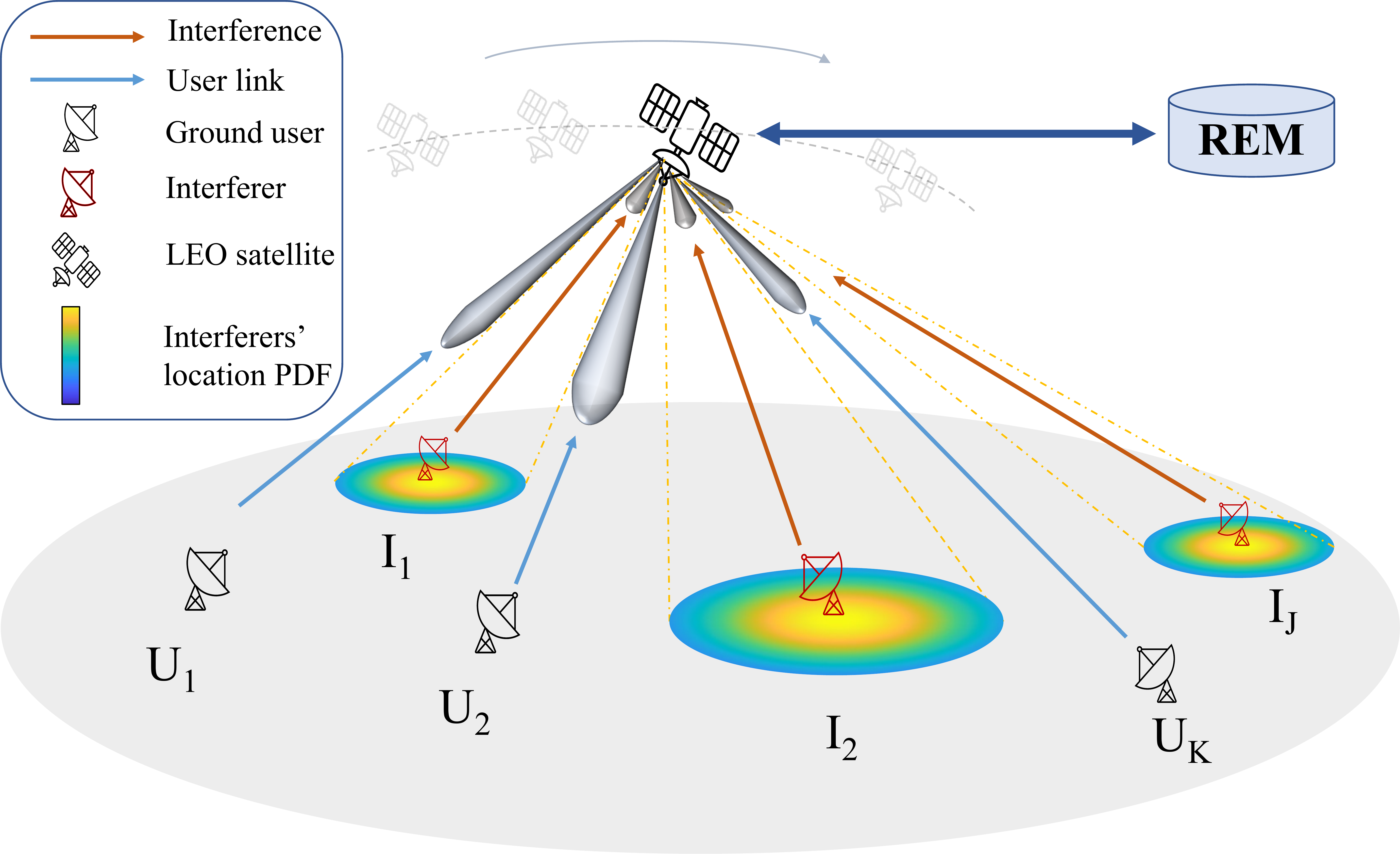}
    \caption{Uplink (ground-to-satellite) transmission scenario in the presence of RFI sources, with uncertainty in their observed locations.}
    \label{fig:Sys_Mod}
\end{figure}

We consider the uplink of a cognitive satellite system, illustrated in Fig.~\ref{fig:Sys_Mod}, where a LEO satellite serves \(K\) ground users in the presence of \(J\) interfering sources. Each interferer is located within a designated uncertainty region. In Fig.~\ref{fig:Sys_Mod}, the interferer's image indicates only its nominal (expected) position. The satellite is assumed to have access to a Radio Environment Map (REM) that provides statistical information on the interferers' locations, obtained through an underlying sensing system. 
\subsection{Beamforming Model}
We consider a LEO satellite equipped with an \(M\times N\) uniform planar antenna array. The total array gain is given by
\begin{equation}
    G(\theta,\phi)=\left|\Xi(\theta,\phi)\Phi(\theta,\phi)\right|^2,
    \label{eq:Array_gain}
\end{equation}
where \(\Xi(\theta,\phi)\) denotes the antenna element factor, which depends on the element type, and \(\Phi(\theta,\phi)\) represents the array factor as a function of the azimuth and elevation angles. 

Each antenna element \(A_{m,n}\), indexed by \(m \in [0, M-1]\) and \(n \in [0, N-1]\), is assigned a complex beamforming weight \(w_{m,n} = a_{m,n} e^{i\psi_{m,n}}\), where \(a_{m,n}\) and \(\psi_{m,n}\) denote the amplitude and phase shift, respectively, used to shape the radiation pattern.  
The collection of beamforming weights is represented in matrix form as \(\mathbf{W} = [w_{m,n}] \in \mathbb{C}^{M \times N}\).  
We neglect mutual coupling effects and assume digital beamforming, which provides higher precision compared to analog beamforming that is limited by discrete amplitude and phase control.  
The inter-element spacings along the \(x\)- and \(y\)-axes are denoted by \(d_x\) and \(d_y\), respectively, and \(\lambda\) denotes the wavelength corresponding to the carrier frequency. Then, the array factor is given by
\begin{equation}
    \Phi(\theta,\phi) = \sum_{m=0}^{M-1} \sum_{n=0}^{N-1} w_{m,n} e^{-i\frac{2\pi}{\lambda}\left(md_x\sin\theta\cos\phi + nd_y\sin\theta\sin\phi\right)}.
    \label{eq:Array_factor}
\end{equation}

Similarly, let \(s_{m,n}\) denote the signal received at element \(A_{m,n}\) from terrestrial users, and define the signal matrix \(\mathbf{S} = [s_{m,n}] \in \mathbb{C}^{M \times N}\).  
Accordingly, the antenna array output \(y\) is expressed as $y = \sum_{m=0}^{M-1} \sum_{n=0}^{N-1} w_{m,n} s_{m,n}$. 
For mathematical convenience, we vectorize the matrices as \(\mathbf{w} = \mathrm{vec}(\mathbf{W})\) and \(\mathbf{s} = \mathrm{vec}(\mathbf{S})\), where \(\mathrm{vec}(\cdot)\) denotes the column-wise vectorization operator.  
Specifically, we have
\(
\mathbf{s}=\left[s_{0,0}, \dotsc, s_{0,N-1}, \dotsc, s_{M-1,N-1}\right]^\mathrm{T} \in \mathbb{C}^{MN\times1},
\) and 
\(
\mathbf{w}=\left[w_{0,0}, \dotsc, w_{0,N-1}, \dotsc, w_{M-1,N-1}\right]^\mathrm{T} \in \mathbb{C}^{MN\times1}
\). Thus, the antenna array output can be compactly rewritten as
\begin{equation}
\label{Ant_arr_out_alt}
    y=\mathbf{w}^H\mathbf{s},
\end{equation}
where \((\cdot)^H\) denotes the Hermitian (conjugate) transpose.

\subsection{Users' and Interferers' Position Model}\label{Subsec_pos_model}

Since ground users are part of the cognitive system, their positions are assumed to be perfectly known at the satellite and are denoted by \(U_k=\left(\theta_k,\phi_k\right)\), 
where \(\theta_k\) and \(\phi_k\) represent the azimuth and off-nadir angles of user \(k\) from the satellite's perspective, respectively, as in~\cite{10622540}.

For interferer \(j\), statistical information on its location, in terms of azimuth and off-nadir angles, is available through the REM. 
Based on this information, a bivariate Gaussian distribution is fitted to model the location uncertainty of each detected interferer \(j\). 
Let \(\mathbf{x}_j=\left(\theta_j,\phi_j\right)^\mathrm{T}\) denote the random vector representing the azimuth and off-nadir angles of interferer \(j\), respectively. 
Then, \(\mathbf{x}_j\) follows a bivariate normal distribution, denoted by \(\mathcal{N}_2(\boldsymbol{\mu}_j,\mathbf{\Sigma}_j)\), with probability density function (PDF)
\begin{equation}
    f_{\mathbf{x}_j}(\theta_j,\phi_j) \!=\! \frac{1}{2\pi \sqrt{|\mathbf{\Sigma}_j|}} \exp\left(\!\!-\!\frac{(\mathbf{x}_j-\boldsymbol{\mu}_j)^\mathrm{T}\mathbf{\Sigma}_j^{-1}(\mathbf{x}_j-\boldsymbol{\mu}_j)}{2}\!\right)
    \label{eq:2_Gaussian}
\end{equation}
where \(\boldsymbol{\mu}_j = \mathbb{E}[\mathbf{x}_j] = \left( \mathbb{E}[\theta_j],\ \mathbb{E}[\phi_j] \right)^\mathrm{T}\) is the mean vector and \(\mathbf{\Sigma}_j = \mathrm{Cov}(\mathbf{x}_j)\) is the covariance matrix. 
Throughout this work, it is assumed that \(\theta_j\) and \(\phi_j\) are uncorrelated, i.e., \(\mathbf{\Sigma}_j\) is a diagonal matrix.

\begin{figure}[t!]
    \centering
    \includegraphics[width=0.5\textwidth]{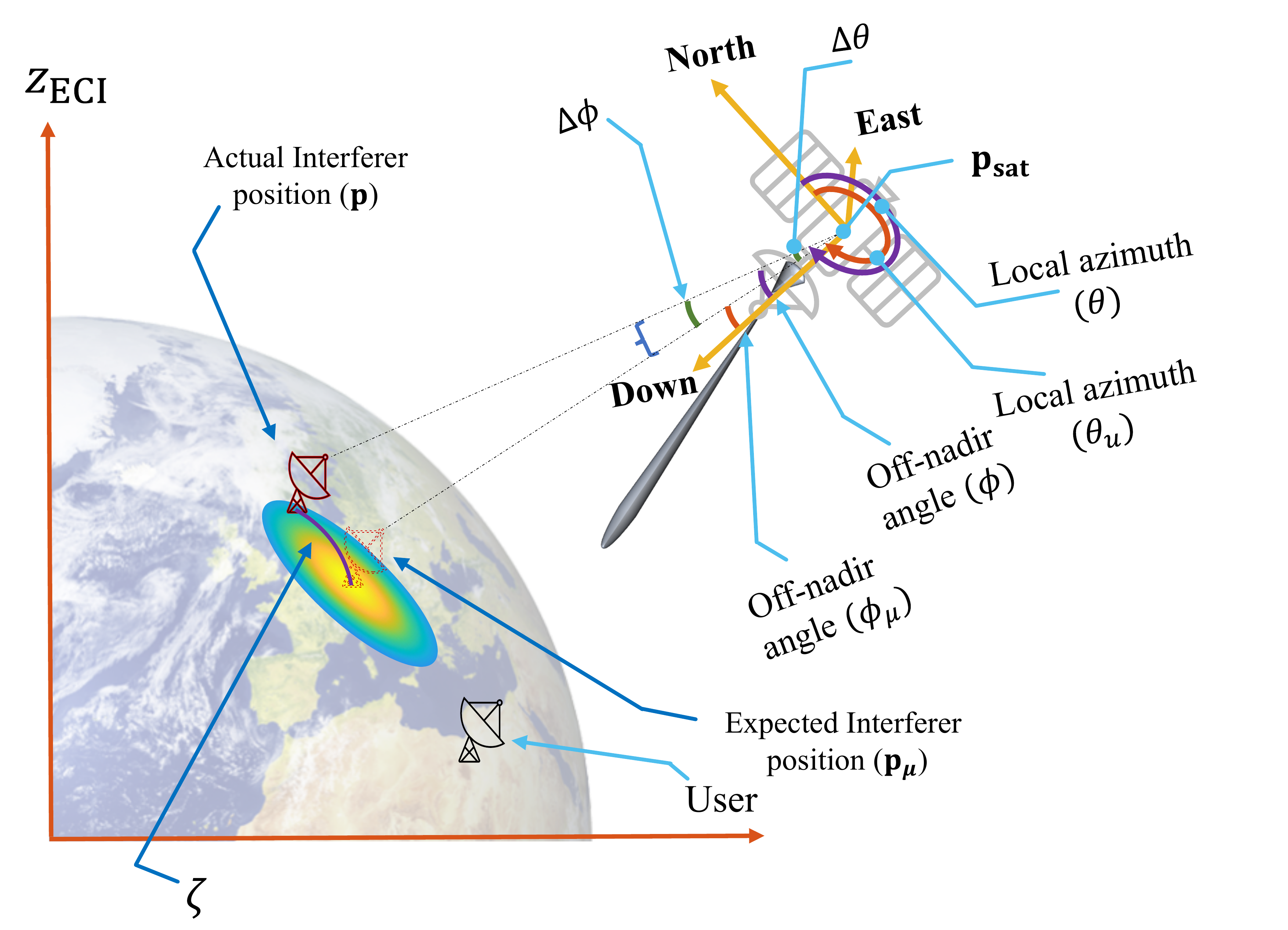}
    \caption{Geometric relationship between the expected and actual interferer positions as observed from a LEO satellite. The orthodromic distance \(\zeta\) quantifies the separation between the two locations.}
    \label{fig:Sat_Beam_Null}
\end{figure}

\subsection{Geometric Analysis}\label{Subsec_Geom_Analysis}

From the satellite's perspective, targets on the Earth's surface are defined in terms of their azimuth (\(\theta\)), elevation (\(\phi\)), and range (\(\rho\)), collectively known as AER coordinates, relative to the satellite position \(\mathbf{p}_\mathrm{sat}\). Fig.~\ref{fig:Sat_Beam_Null} illustrates this concept, showing how the location of an interference source is observed from space.

The expected location of an interferer in AER coordinates is denoted by \(\mathbf{p}_\mu^\mathrm{AER} = (\theta_\mu, \phi_\mu, \rho_\mu)\), while its actual location is \(\mathbf{p}^\mathrm{AER} = (\theta, \phi, \rho)\), lying within the uncertainty region defined by the corresponding PDF. A mismatch between \(\mathbf{p}_\mu^\mathrm{AER}\) and \(\mathbf{p}^\mathrm{AER}\) induces angular deviations \((\Delta\theta, \Delta\phi)\), which in turn correspond to an orthodromic distance \(\zeta\) over the Earth's surface between the two locations.

Understanding the relationship between the angular differences \((\Delta\theta, \Delta\phi)\) and the resulting orthodromic distance \(\zeta\) is crucial for the accurate design of the satellite's spatial nulling regions. To transform AER coordinates into geodetic coordinates (longitude, latitude, and altitude), the AER coordinates are first mapped into the local North-East-Down (NED) frame, denoted by \(\mathbf{p}^\mathrm{NED} = [N, E, D]^\mathrm{T}\). The mapping follows the transformations as $X = \rho \cos(\phi) \sin(\theta);\, Y= \rho \cos(\phi) \cos(\theta);\, Z = -\rho \sin(\phi)$, where \(N = X\), \(E = Y\), and \(D = -Z\). 
Then, we convert from NED coordinates to Earth-Centered, Earth-Fixed (ECEF) coordinates using
\begin{equation}
    \mathbf{p}^\mathrm{ECEF} = \mathbf{p}^\mathrm{ECEF}_\mathrm{sat} + \mathbf{R}^\mathrm{N\!-\!E} \mathbf{p}^\mathrm{NED},
    \label{eq:ECEF_coordinates}
\end{equation}
where \(\mathbf{p}^\mathrm{ECEF}_\mathrm{sat} = [X_\mathrm{sat},\ Y_\mathrm{sat},\ Z_\mathrm{sat}]^\mathrm{T}\) denotes the satellite's ECEF position. These coordinates are obtained from the satellite's geodetic coordinates \(\mathbf{p}^\mathrm{GEO}_\mathrm{sat} = (\Lambda_s,\ \Theta_s,\ h_s)\), where \(\Lambda_s\), \(\Theta_s\), and \(h_s\) represent its longitude, latitude, and altitude, respectively, according to
\begin{subequations}\label{eq:ECEF-sat}
\begin{align}
X_\mathrm{sat} &= (R_\mathrm{N} + h_s) \cos(\Theta_s) \cos(\Lambda_s)\\
Y_\mathrm{sat} &= (R_\mathrm{N} + h_s) \cos(\Theta_s) \sin(\Lambda_s)\\
Z_\mathrm{sat} &= \left( \frac{b^2}{a^2} R_\mathrm{N} + h_s \right) \sin(\Theta_s)
\end{align}
\end{subequations}
where \(R_\mathrm{N}\) is the prime vertical radius of curvature, given as 
\begin{equation}
    R_\mathrm{N} = \frac{a}{\sqrt{1 - e^2 \sin^2(\Theta_s)}},
    \label{eq:RN}
\end{equation}
with \(a\) and \(b\) denoting the Earth's semi-major and semi-minor axes, respectively, and \(e\) representing its eccentricity.

The rotation matrix \(\mathbf{R}^\mathrm{N\!-\!E}\) in~\eqref{eq:ECEF_coordinates} transforms vectors from the NED frame to the ECEF frame. It is given by
\begin{equation}
    \mathbf{R}^\mathrm{N\!-\!E} \!=\! 
    \begin{bmatrix}
    -\sin(\Lambda_s) & -\sin(\Theta_s)\cos(\Lambda_s) & \cos(\Theta_s)\cos(\Lambda_s)\\
    \cos(\Lambda_s) & -\sin(\Theta_s)\sin(\Lambda_s) & \cos(\Theta_s)\sin(\Lambda_s)\\
    0 & \cos(\Theta_s) & \sin(\Theta_s)
    \end{bmatrix}
    \label{eq:rotationNEDtoECEF}
\end{equation}
where the ECEF X-axis points towards the intersection of the Equator and the Greenwich meridian, the Y-axis points towards the intersection of the Equator and \(90^\circ\) east longitude, and the Z-axis points towards the North Pole. 

Finally, \(\mathbf{p}^\mathrm{ECEF}\) is converted to geodetic coordinates using
\begin{subequations}\label{eq:ECEF-Geodetic}
\begin{align}
\Lambda &= \arctan2(X, Y)\\
\Theta &= \arctan2\left(Z, \sqrt{X^2 + Y^2}\right)\\
h &= \frac{\sqrt{X^2 + Y^2}}{\cos(\Theta)} - R_\mathrm{N}.
\end{align}
\end{subequations}
Here, \(\Lambda\) and \(\Theta\) denote the longitude and latitude, respectively, and \(h\) represents the altitude. The latitude \(\Theta\) can be further refined iteratively using
\[
\Theta \leftarrow \arctan \,\, 2\left(Z + R_\mathrm{N} e^2 \sin(\Theta), \sqrt{X^2 + Y^2}\right),
\]
with the corresponding update of \(h\) following each iteration.

Once the geodetic coordinates of the expected interferer location \(\mathbf{p}^\mathrm{GEO}_\mu = (\Lambda_\mu, \Theta_\mu, h_\mu)\) and its actual location \(\mathbf{p}^\mathrm{GEO} = (\Lambda, \Theta, h)\) are obtained, the Haversine formula is used to calculate the orthodromic (great-circle) distance \(\zeta\) between them. The angular distance \(\varphi\) and the final distance \(\zeta\) are computed as follows:
\begin{subequations}\label{eq:Haversine_distance}
\begin{align}
\eta &= \sin^2\left(\frac{\Delta \Theta}{2}\right) + \cos(\Theta_\mu) \cos(\Theta) \sin^2\left(\frac{\Delta \Lambda}{2}\right) \label{eq:Haversine_distance_a}\\
\varphi &= 2 \arctan2\left(\sqrt{\eta}, \sqrt{1-\eta}\right) \label{eq:Haversine_distance_b}\\
\zeta &= R_\mathrm{e} \cdot \varphi \label{eq:Haversine_distance_c}
\end{align}
\end{subequations}
where \(R_\mathrm{e}\) is the Earth's mean radius, \(\Delta \Theta = \Theta - \Theta_\mu\) and \(\Delta \Lambda = \Lambda - \Lambda_\mu\) are the differences in latitude and longitude, respectively.

\begin{figure}[t!]
     \centering
         \subfloat[\(\Delta\theta=0.5^\circ\)]{\label{fig:Arc_DeltaPhi_DeltaTheta_a}\includegraphics[width=0.5\textwidth]{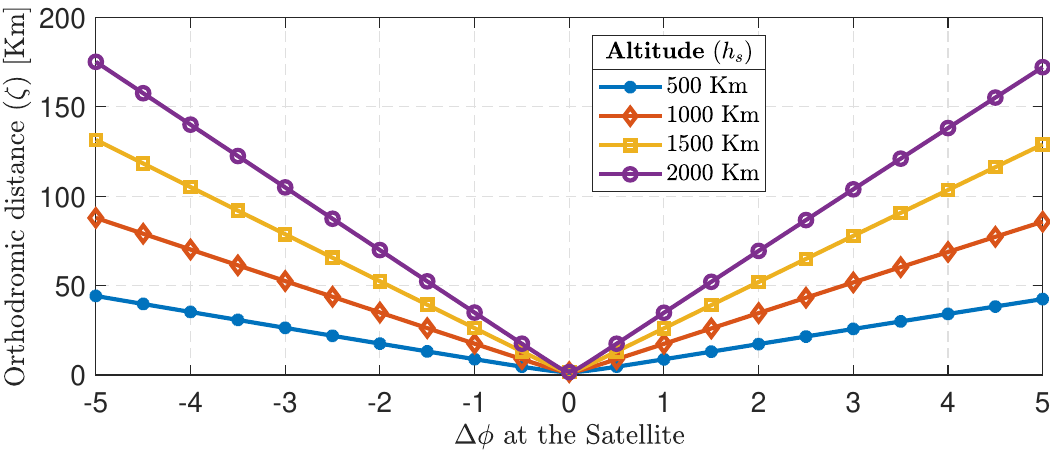}}\\
         \vspace{-4mm}
         \subfloat[\(\Delta\phi=0.5^\circ\)]{\label{fig:Arc_DeltaPhi_DeltaTheta_b}\includegraphics[width=0.5\textwidth]{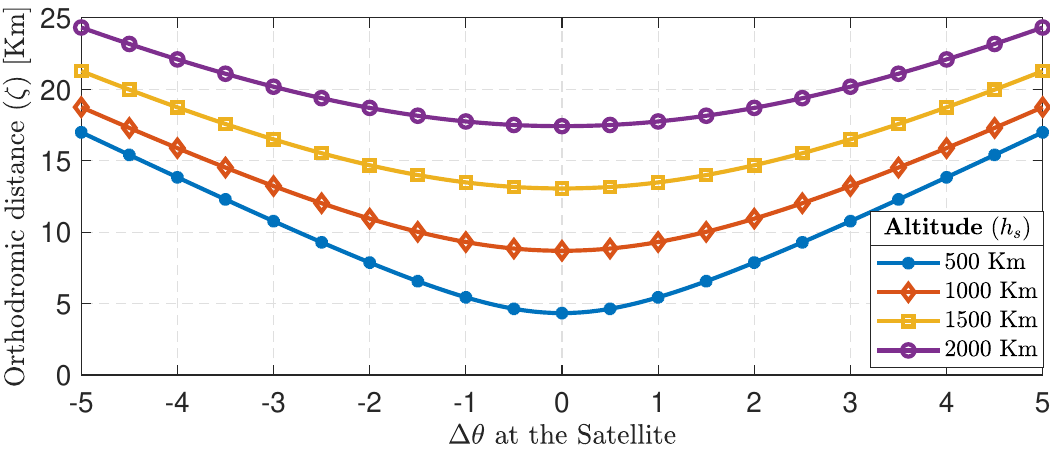}}
       \caption{Orthodromic distance \(\zeta\) on the Earth's surface between the expected and actual interferer locations, observed from a LEO satellite at various altitudes.}
    \label{fig:Arc_DeltaPhi_DeltaTheta}
\end{figure}

In Fig.~\ref{fig:Arc_DeltaPhi_DeltaTheta}, the orthodromic distance \(\zeta\) (in kilometers) between the expected and actual target locations is presented for different satellite altitudes. The satellite is positioned at \(\Lambda_s = 138.53^\circ\) and \(\Theta_s = -22.024^\circ\). In Fig.~\ref{fig:Arc_DeltaPhi_DeltaTheta_a}, the latitude deviation is fixed at \(\Delta\theta = 0.5^\circ\), while in Fig.~\ref{fig:Arc_DeltaPhi_DeltaTheta_b}, the longitude deviation is fixed at \(\Delta\phi = 0.5^\circ\). As observed, the distance between the two points increases with the satellite's altitude. Even small deviations in \(\Delta\theta\) and \(\Delta\phi\) correspond to several kilometers on the Earth's surface, which is significant enough to critically impact interference mitigation strategies based on beamforming and nulling designs if such uncertainties are not considered. This underscores the necessity of high-resolution beamforming and nulling to effectively maintain system performance.

\section{Null-Shaping for Interference Mitigation}\label{Sec_Null-shaping}
To enhance classical beamforming-based interference mitigation, we propose a null-shaping framework that improves robustness against imperfect knowledge of RFI source locations. Our approach incorporates the probability of interferer \(j\) being at a given angular position \((\theta_j,\ \phi_j)^T\) as a weighting factor within the beamforming optimization process. For each interferer \(j\), once \(\boldsymbol{\mu}_j\) and \(\boldsymbol{\Sigma}_j\) are determined as described in Section~\ref{Subsec_pos_model}, its probability distribution is sampled to extract relevant information for null design. Specifically, the expected angular positions \(\mu_{\theta_j}\), \(\mu_{\phi_j}\) and the corresponding standard deviations \(\sigma_{\theta_j}\), \(\sigma_{\phi_j}\) are used to generate a set of angular positions for shaping the nulls.

\subsection{Probabilistic Null Design with Angular Uncertainty}
We select \(L\) samples evenly spaced within the ranges \(\left[\mu_{\theta_j}-\kappa\sigma_{\theta_j},\,\mu_{\theta_j}+\kappa\sigma_{\theta_j}\right]\) and \(\left[\mu_{\phi_j}-\kappa\sigma_{\phi_j},\,\mu_{\phi_j}+\kappa\sigma_{\phi_j}\right]\), defining the most likely angular positions \(\boldsymbol{\theta}_j\) and \(\boldsymbol{\phi}_j\) to be considered in the null-shaping design. The parameter \(\kappa\in\mathbb{Z}^+\) controls the range by specifying the number of standard deviations included in the beamforming adaptation. Without loss of generality, expressions for \(\boldsymbol{\theta}_j\) and \(\boldsymbol{\phi}_j\) are given in \eqref{eq:theta_range} and \eqref{eq:phi_range} for the case \(L=3\) and \(\kappa=1\) as
\begin{equation}
    \boldsymbol{\theta}_j = (\mu_{\theta_j} - \sigma_{\theta_j},\ \mu_{\theta_j},\ \mu_{\theta_j} + \sigma_{\theta_j}),
    \label{eq:theta_range}
\end{equation}
\begin{equation}
    \boldsymbol{\phi}_j = (\mu_{\phi_j} - \sigma_{\phi_j},\ \mu_{\phi_j},\ \mu_{\phi_j} + \sigma_{\phi_j}).
    \label{eq:phi_range}
\end{equation}
The Cartesian product of \(\boldsymbol{\theta}_j\) and \(\boldsymbol{\phi}_j\) defines the set of angular pairs collected in the matrix \(\mathbf{I}_j \in \mathbb{R}^{L^2 \times 2}\). 
For each angle pair in \(\mathbf{I}_j\), we compute the corresponding probability using \eqref{eq:2_Gaussian} to form the vector \(\mathbf{p}_j = [p_1\ p_2\ \cdots\ p_{L^2}]^T \in \mathbb{R}^{L^2 \times 1}\).
The probability vector \(\mathbf{p}_j\) acts as a weighting factor, emphasizing the contribution of each sampled direction in the null-shaping optimization. 

Accordingly, the beamforming gain toward interferer \(j\), denoted \(G_j(\boldsymbol{\theta}_j,\boldsymbol{\phi}_j)\), is defined as the weighted sum of the array gains at sampled angular positions, given by 
\begin{align}
    G_j\left(\boldsymbol{\theta}_j,\boldsymbol{\phi}_j\right) 
    &= \sum_{z=1}^{L^2}\, p_j^z G\left(\theta_j^z,\phi_j^z\right)
    = \sum_{z=1}^{L^2}\, p_j^z \left|\Phi\left(\theta_j^z,\phi_j^z\right) \right|^2,
    \label{eq:Inter_j_gain}    
\end{align}
where \(G(\theta,\phi)\) is given by \eqref{eq:Array_gain}, and the antenna element pattern \(\Xi(\theta,\phi)\) is assumed to be omni-directional, i.e., \(\Xi(\theta,\phi) = 1\), for simplicity and practical relevance.

\subsection{Beamforming Metric and Optimization Problem}
With the angular sampling and associated weights defined, we now formulate the beamforming optimization objective. The goal is to maximize the array gain toward the intended users while minimizing it toward radio frequency interference (RFI) sources. These competing objectives are jointly captured by the {\it mitigation effectiveness} metric, denoted by \(\Psi(\cdots)\), and defined as
\begin{align}
     \Psi \!\!=\!\! \frac{\frac{1}{K} \sum_{k=1}^{K} G_k(\theta_k,\phi_k)}{\frac{1}{J} \sum_{j=1}^{J} G_j(\boldsymbol{\theta}_j, \boldsymbol{\phi}_j)} 
     \!=\! \frac{\frac{1}{K} \sum_{k=1}^{K} \left|\Phi(\theta_k,\phi_k)\right|^2}{\frac{1}{J} \sum_{j=1}^{J} \sum_{z=1}^{L^2} p_j^z \left|\Phi\left(\theta_j^z,\phi_j^z\right)\right|^2}
     \label{eq:Psi}
\end{align}
 which quantifies the {\it ratio of the average beamforming gain toward the users to that toward the interferers}. Here, \(p_j^z\) denotes the sampling weight associated with the \(z\)th angular perturbation around the \(j\)th interferer, reflecting the relative likelihood of the spatial offset due to location uncertainty.

Our objective is to determine the optimal beamforming weight vector \(\mathbf{w} = [w_{m,n}] \in \mathbb{C}^{MN \times 1}\) that maximizes \(\Psi\left(\{\theta_k, \phi_k\}_{k=1}^{K}, \{\theta_j^z, \phi_j^z, p_j^z\}_{j=1,z=1}^{J,L^2}\right)\), while satisfying practical design constraints, including a normalization constraint   $\|\mathbf{w}\|_2^2 \leq 1$ and per-element phase constraints \(\arg(w_{m,n}) \in [0,2\pi)\).  
Accordingly, the beamforming optimization problem is formulated as
\begin{align}
\underset{\mathbf{w}}{\text{maximize}} \quad & \Psi\left(\{\theta_k, \phi_k\}_{k=1}^{K}, \{\theta_j^z, \phi_j^z, p_j^z\}_{j=1, z=1}^{J, L^2}\right) \\
\text{subject to} \quad & \|\mathbf{w}\|_2^2 \leq 1 \label{eq:power_constraint} \\
& 0 \leq \arg(w_{m,n}) < 2\pi, \quad \forall m,n. \label{eq:phase_constraints}
\end{align}

\subsection{Heuristic Optimization Approach}
Due to the nonconvex and nonlinear nature of the objective function \(\Psi(\cdot)\) with respect to the beamforming vector \(\mathbf{w}\), finding a closed-form global optimum is generally intractable. The nonconvexity arises from the fractional structure of \(\Psi\) and the per-element phase constraints on \(\mathbf{w}\). Furthermore, the computational complexity grows with the number of users $K$, interferers $J$, and antenna elements $MN$, as well as with the granularity of the angular sampling over the interference regions.   

To obtain a practical solution, we employ a heuristic optimization strategy based on Particle Swarm Optimization (PSO) and Intermediate Point (IP) methods, which offer a favorable tradeoff between computational efficiency and beamforming performance, as demonstrated in prior works such as \cite{10622540}. These algorithms are well-suited for handling the highly nonconvex search space introduced by the physical constraints on  phase and power. Although more sophisticated methods, such as semidefinite relaxation (SDR) or manifold optimization, could be considered, they often incur significantly higher computational overhead, especially for large-scale arrays. In this work, our focus remains on achieving effective RFI mitigation under practical constraints with manageable complexity.

\begin{figure}[h]
    \centering
    \includegraphics[width=0.5\textwidth]{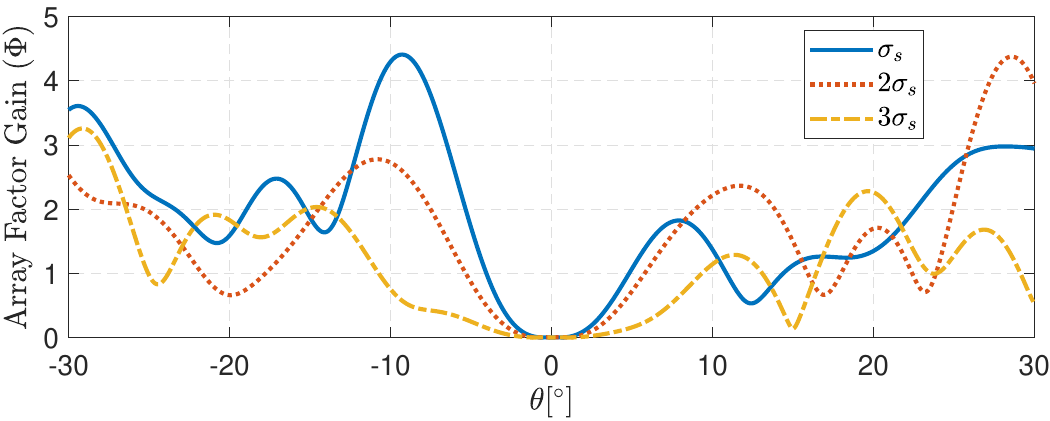}
    \caption{Antenna array's radiation pattern, showing nulling design at \(\mu_\theta=0^\circ\) considering uncertainty within different number of \(\sigma_s\) values, when \(\sigma_s=\pm1\).}
    \label{fig:Nulling_sigmas}
\end{figure}

\section{Numerical Results}\label{Sec_Results}

Numerical simulations consider a LEO satellite at 800~km altitude operating in the \(\mathrm{K_a}\)-band. The satellite follows a near-circular orbit (\(e \approx 0\)) with a semi-major axis of 7173~km, inclination of \(86.39^\circ\), right ascension of ascending node (RAAN) \(146.16^\circ\), argument of periapsis \(269.5^\circ\), and true anomaly \(0.6^\circ\).
\begin{figure}[t!]
    \centering
    \subfloat[Beamforming mitigation with \(\sigma_s=0^\circ\)]{
        \includegraphics[width=0.93\columnwidth]{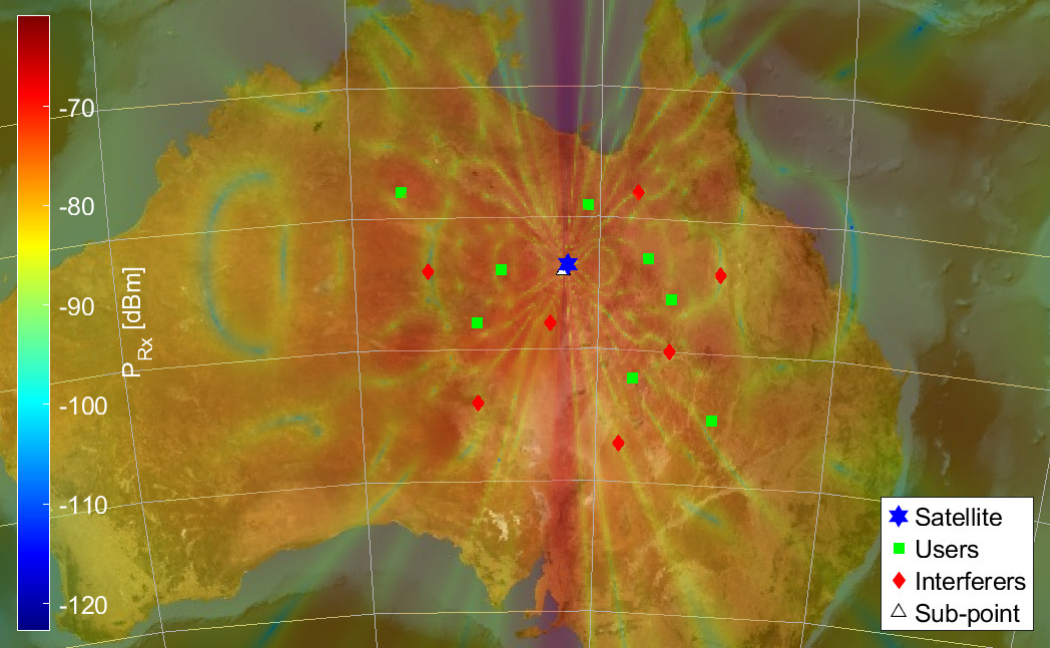}
        \label{fig:3D_radiation_patterns_a}
    }

    \vspace{2mm} 

    \subfloat[Beamforming mitigation with \(\sigma_s=2^\circ\)]{
        \includegraphics[width=0.93\columnwidth]{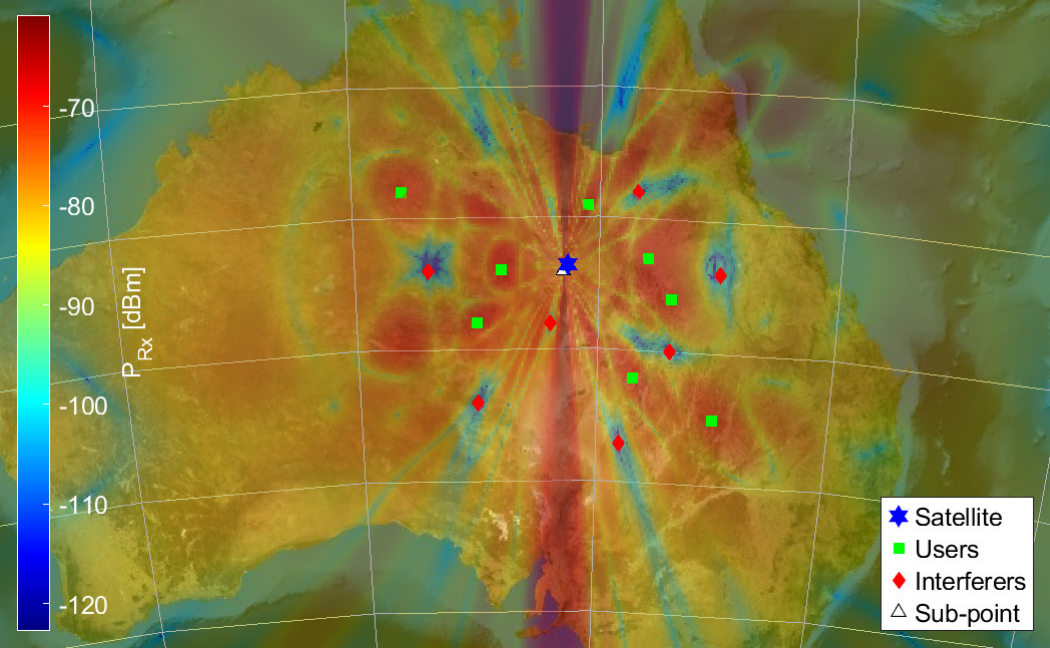}
        \label{fig:3D_radiation_patterns_b}
    }

    \caption{Received power levels over northern Australia observed by a \(24\times24\) planar array at the satellite.}
    \label{fig:3D_radiation_patterns}
\end{figure}

In Fig.~\ref{fig:Nulling_sigmas}, we demonstrate the nulling capability of a \(1 \times 20\) linear array targeting an interferer at \(\mu_\theta = 0^\circ\). The null design incorporates angular uncertainty modeled via standard deviations \(\sigma_s = 1^\circ\), with beam patterns synthesized for \(\sigma_s\), \(2\sigma_s\), and \(3\sigma_s\). As expected, increasing the uncertainty range leads to progressively wider nulls, illustrating the flexibility of the proposed algorithm in adapting to varying design criteria.

\begin{figure}[t!]
    \centering
    \includegraphics[width=0.5\textwidth]{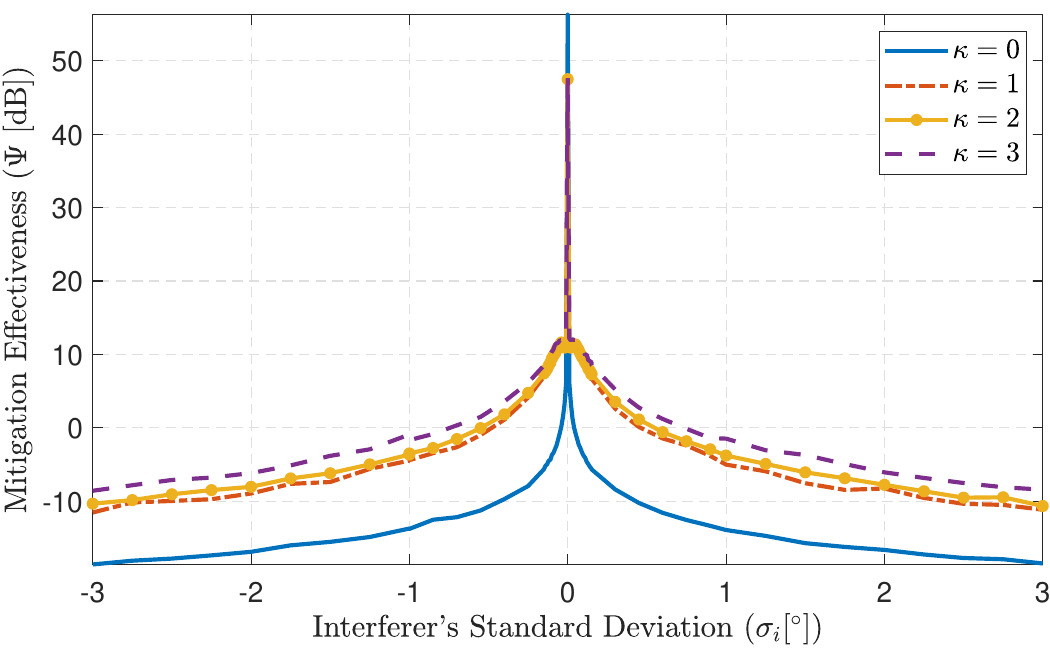}
    \caption{Mitigation Effectiveness for an \(8\times8\) planar array vs Interferer's standard deviation for various \(\kappa\) values and \(\sigma_{s}=0.1^\circ\) for the null-shaping design.}
    \label{fig:Mitigation_kappas}
\end{figure}

\begin{figure}
    \centering
    \includegraphics[width=0.5\textwidth]{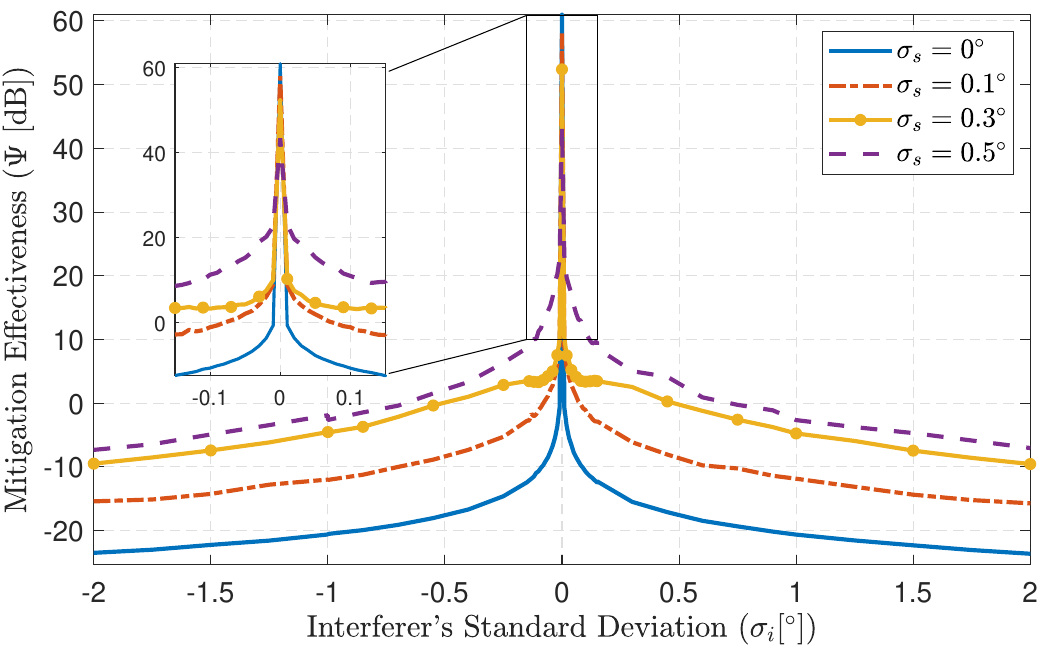}
    \caption{Mitigation Effectiveness for an \(8\times8\) planar array vs Interferer's standard deviation for various \(\sigma_s\) values and \(\kappa=1\) for the null-shaping design.}
    \label{fig:Mitigation_sigmas}
\end{figure}

Fig.~\ref{fig:3D_radiation_patterns} compares beam patterns generated by a \(24 \times 24\) planar array under perfect interferer location knowledge (\(\sigma_s = 0^\circ\)) and under angular uncertainty (\(\sigma_s = 2^\circ\)) for a multiple-user and multiple-interferer scenario. In both cases, the expected positions of the users and interferers are identical. As shown in Fig.~\ref{fig:3D_radiation_patterns_a}, the nulls are sharp and precisely aligned with the interferers when no uncertainty is assumed. In contrast, Fig.~\ref{fig:3D_radiation_patterns_b} shows broader nulls resulting from the assumed uncertainty. Together with Fig.~\ref{fig:Nulling_sigmas}, these results illustrate the proposed algorithm’s effectiveness in shaping the null space according to interference uncertainty.

Fig.~\ref{fig:Mitigation_kappas} shows the mitigation effectiveness of an \(8 \times 8\) planar array under varying interferer standard deviations \(\sigma_i\), for a fixed null-shaping for \(\sigma_s = 0.1^\circ\) and three \(\kappa\) values. Mitigation effectiveness quantifies the relative attenuation of interference compared to the desired signal. The \(\kappa = 0\) case (blue curve) represents a sharp null assuming perfect knowledge of the interferer’s direction; while highly effective at \(\sigma_i = 0^\circ\), performance degrades rapidly as \(\sigma_i\) increases. In contrast, higher \(\kappa\) values broaden the null to accommodate uncertainty, maintaining more robust interference suppression across a wider range of \(\sigma_i\). These results underscore the trade-off between null sharpness and robustness in uncertain environments.

Fig.~\ref{fig:Mitigation_sigmas} shows how mitigation effectiveness varies with the interferer’s angular uncertainty \(\sigma_i\), for different null-shaping values \(\sigma_s\) at the satellite (\(\kappa=1\)). The blue curve (\(\sigma_s=0^\circ\)) yields the highest attenuation at \(\sigma_i=0^\circ\), but its performance degrades rapidly with even minimal deviation, due to the narrow null. In contrast, the red, yellow, and purple curves (\(\sigma_s=0.1^\circ\), \(0.3^\circ\), and \(0.5^\circ\), respectively) exhibit broader and more resilient nulls, maintaining higher suppression across a wider \(\sigma_i\) range. These results highlight the trade-off between peak suppression and robustness to directional uncertainty, governed by the choice of \(\sigma_s\). 

\begin{figure}
    \centering
    \includegraphics[width=0.5\textwidth]{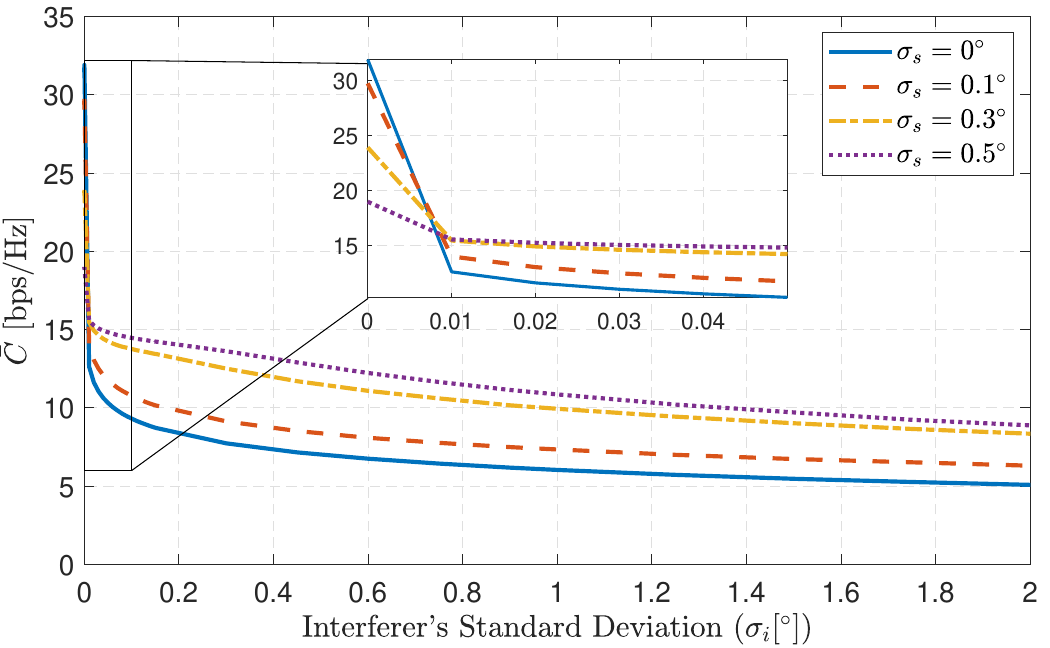}
    \caption{Capacity at the satellite for an \(8\times8\) planar array vs Interferer's standard deviation for various \(\sigma_s\) values and \(\kappa=1\) for the null-shaping design.}
    \label{fig:Capacity_sigmas}
\end{figure}
Fig.~\ref{fig:Capacity_sigmas} illustrates the achievable capacity [bps/Hz] at the satellite under null-shaping beamforming, with a user at \((\Lambda=136^\circ, \Theta=-22^\circ)\) and an interferer at \((\Lambda=141.5^\circ, \Theta=-19^\circ)\). The \(\sigma_s=0^\circ\) case achieves the highest capacity when the interferer is exactly at the expected location, but suffers a sharp degradation with slight mismatch. In contrast, \(\sigma_s>0^\circ\) designs provide more robust capacity in the presence of location uncertainty by broadening the null, trading off peak performance for improved reliability—crucial when precise interferer localization is impractical.

\section{Conclusions}\label{Sec_Conclusions}
This work addressed the challenge of uplink interference mitigation in LEO satellite systems under uncertainty in the location of terrestrial RFI sources. We proposed a robust beamforming framework using a planar antenna array at the satellite, where null shaping is guided by a weighted optimization scheme that incorporates statistical information about RFI location uncertainty. The proposed method enables adaptive adjustment of the radiation pattern to improve interference suppression in the presence of ambiguous location data, enabling balance between robustness and efficiency that is beyond the capability of purely deterministic methods. Simulation results demonstrated that significant performance gains can be achieved by tailoring the nulling region based on the probability distribution of interferer positions, offering enhanced robustness compared to deterministic designs. Future work will investigate the trade-offs between array size and achievable nulling resolution, and extend the proposed framework to dynamic beamwidth control under satellite motion, with particular attention to implementation challenges arising from satellite payload processing constraints. In addition, to assess practical applicability within a cognitive satellite system, we will investigate REM-induced errors and their impact on algorithm performance. 


\bibliographystyle{IEEEtran}

\end{document}